\documentclass[english,notitlepage,12pt]{article}
\usepackage[T1]{fontenc}
\usepackage[utf8]{inputenc}
\usepackage{amssymb}
\usepackage{amsmath}
\usepackage{graphicx}
\usepackage{esint}
\usepackage{xcolor}
\usepackage{hyperref}
\usepackage{babel}
\usepackage{authblk}
\usepackage{abstract}
\usepackage{caption}
\usepackage{perpage}
\usepackage{footmisc}

\newcommand{\GeV}{\mbox{GeV}} 
\newcommand{\Br}{\mathrm{Br}}

\newcommand{\Th}{\mathrm{th}}

\addto\captionsenglish{}
\addto\captionsenglish{}

\definecolor{urlcolor}{RGB}{15,25,112}
\hypersetup{
    bookmarks=true,         
    unicode=false,          
    pdftoolbar=true,        
    pdfmenubar=true,        
    pdffitwindow=false,     
    pdfstartview={FitH},    
    pdftitle={LHC-chib},    
    pdfauthor= {Stanislav Poslavsky},     
    pdfsubject={LHC-chib}, 
    pdfcreator={Stanislav Poslavsky},
    pdfproducer={hz}, 
    pdfkeywords={keyword1} {key2} {key3}, 
    pdfnewwindow=true,      
    colorlinks=true,       
    linkcolor=black,          
    citecolor=black,        
    filecolor=magenta,      
    urlcolor=urlcolor           
}

\MakePerPage[2]{footnote}

\begin{document}

\title{\textbf{Production of $\chi_{b}$-mesons at LHC}}
\author[1]{A.K. Likhoded\thanks{Anatolii.Likhoded@ihep.ru}}
\author[1]{A.V. Luchinsky\thanks{Alexey.Luchinsky@ihep.ru}}
\author[1]{S.V. Poslavsky\thanks{stvlpos@mail.ru}}
\affil[1]{\slshape{Institute for High Energy Physics, Protvino, Russia}}
\date{}
\maketitle

\renewcommand{\abstractname}{ }
\renewcommand{\abstractnamefont}{\normalfont\bfseries}
\renewcommand{\abstracttextfont}{\normalfont}
\newcommand{\diff}[2]{\frac{\mathrm{d} #1}{\mathrm{d} #2}}

\vspace{-1.0cm}
\begin{abstract}
\vspace{-0.5cm}
In the present paper we discuss the $P$-wave bottomonium production within both
color octet and color singlet models in NLO at LHC energies. We calculate
the cross section and the transverse momentum distributions for the
$\chi_{1,2b}$ states. We obtain, that the ratios of bottomonium and charmonium
spin states production are fundamentally complementary at different $p_T$
scales. We give predictions for the ratio of the $n=2$ and $n=1$ radial
excitations production cross sections.\\[12pt]
\end{abstract}

\section{Introduction}

Studies of heavy quarkonia give a deeper understanding
of the Quantum Chromodynamics. In the recent measurements made by
ATLAS \cite{Aad:2011ih} and D0 \cite{D0:2012gh} Collaborations the new data for
the $b\bar{b}$ systems was obtained. The radial excitation of the $P$-wave
$\chi_{b}$ states was observed in radiative transitions to the $S$-wave
$\Upsilon$ states, while $\chi_{b}(3P)$ was seen for the first
time. The $\chi_b$ system is a triplet of closely spaced states with total spin
$J=0$, 1 and 2, that were not observed separately at ATLAS yet.
Usually,  only $1^{++}$ and $2^{++}$ can be detected, since 
$0^{++}$ state has very small radiative branching fraction.
So, in the present paper, we shall focus on a production of $1^{++}$ and
$2^{++}$ charmonium and bottomonium states. We shall consider the production
with high transverse momentum of the final quarkonium. 

Main contribution to the production processes at high energies ($\sim$ TeV)
is given by gluon-gluon subprocesses. The problem of the quarkonia production
can be divided into two parts. The first part of the problem is to obtain the
nonzero transverse momentum of the final quarkonia using the integrated partonic
distributions (PDFs). The second part consists in the hadronization process,
i.e.
formation of the quarkonia with certain quantum numbers. For example, let us
consider the production of the $|^3S_1\rangle$ ($1^{--}$) state. The colorless
state with such quantum numbers cannot be produced from two gluons because of
the charge parity conservation. So, the production of the color singlet state
with the additional emission of a gluon in the final state was introduced
\cite{Kartvelishvili:1978id,Gershtein:1980jb} to explain the quarkonia
production. Additional
emission of the gluon gives the $p_T$-distribution and removes the prohibition
of the $1^{--}$ state production. It is well known, that such a model, in its
naive understanding, is in the contradiction with the experimental data at high
$p_T$. In the case of $P$-wave meson production the situation is different:
$0^{++}$ and $2^{++}$ states can be produced from two gluons, but in the
collinear approximation (with integrated over $p_T$ partonic distributions) the
transverse momentum distribution of final quarkonium cannot be obtained.
Moreover, $|^3P_1\rangle$ state cannot be produced due to Landau-Yang theorem,
which forbids the production of axial meson from two massless gluons. These
problems have led to the approach, known as $k_{T}$-factorization
\cite{Gribov:1984tu,Catani:1990eg,Collins:1991ty,Kniehl:2006sk}, in which
unintegrated partonic distributions are used. However, we think that such an
approach should play a significant role at the low $p_T$ region only, while in
the high $p_T$ region the dominant contribution to the $p_T$ is given by the
processes with emission of a single hard gluon in the final state.  

In this paper we try to resolve the above problems considering a set of
diagrams, in which high $p_T$ is achieved by the emission of single hard gluon,
mostly from the initial state. It turns out that all three $P$-waves states,
including $1^{++}$ are allowed. As we shall see, the color-singlet contributions
give $p_T$ dependence, which is similar to the experiment data for $\chi_c$
mesons, for example from CDF \cite{Abe:1997yz}, but the absolute normalization
of the cross section is several times smaller than in the experiment. To
resolve this contradiction we took into account the color-octet contributions
to the cross sections. In the next section we show that if we consider the
color-octet matrix elements as free parameters, the good agreement with the
experimental data can be obtained for the case of $\chi_c$, and the predictions
for the $\chi_b$ states can be provided. Next, we consider the cross section
ratio of the $2^{++}$ and $1^{++}$ production both for $\chi_c$ and $\chi_b$,
which was recently measured for charmed mesons at the LHCb detector
\cite{LHCb:2012mr}, and show, that our approach gives the value, lying at the
lower limit of uncertainties of the LHCb data. There is no such experimental
data on the bottomonium production yet and our result should be considered as
the theoretical predictions for further experiments. Next, we consider the ratio
of the $n=2$ and $n=1$ radial excitations, which is not measured yet and present
our theoretical predictions.

\section{$\chi_c$ production at high $p_T$}

The cross section of heavy quarkonium production in hadronic interaction can be
expressed through the cross sections of hard subprocesses:
\begin{eqnarray}
\label{eq:sigma}
d\sigma\left[pp\to\mathcal{Q}+X\right] & = & \int
dx_{1}dx_{2}f_{g}\left(x_{1};Q\right)f_{g}\left(x_{2};Q\right)d\hat{\sigma}%
[\mathcal{Q}]
\end{eqnarray}
Here, $\mathcal{Q}$ is one of the $|^3P_J\rangle$ states, $x_{1,2}$ are the
momentum fractions of the partons, $f_{g}\left(x_{1,2};Q\right)$ are the gluon
distribution functions in initial hadrons, and
$d\hat{\sigma}\left[\mathcal{Q}\right]$ is the cross section of the meson
production at the partonic level.

\begin{figure}[h]
\begin{centering}
\includegraphics[width=15cm]{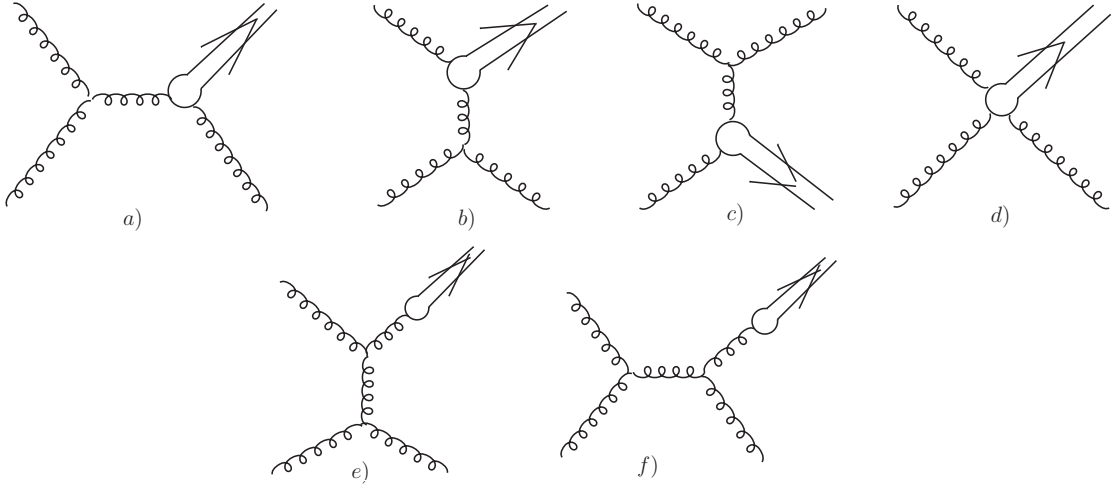}
\par\end{centering}
\caption{The Feynman diagrams of the $gg\to\chi_{b}g$ subprocesses. The first
four diagrams are valid both for color-singlet and color-octet mechanism, while
the last two diagrams corresponds to the octet production only.}
\label{fig:diags}
\end{figure}

As already mentioned in the Introduction, in order to obtain nonzero transverse
momentum of final quarkonia using the usual partonic distributions, it is
necessary to consider next to leading order subprocesses with the emission of
an additional gluon, i.e. $gg\to\chi_{bJ}g$. The corresponding Feynman diagrams
are shown in Fig.~\ref{fig:diags}. The partonic cross sections of these
reactions were calculated by a number of authors
\cite{Gastmans:1987be,Cho:1995ce,Klasen:2003zn,Meijer:2007eb}, but result
presented in \cite{Klasen:2003zn} disagrees with others works. The reason is
that diagrams shown in Fig.\ref{fig:diags} include 3-gluon vertex and the
additional ghost contributions should be included. Our calculations
performed in QCD axial gauge (which does not require additional ghosts
contributions) reproduces \cite{Gastmans:1987be,Cho:1995ce,Meijer:2007eb}
results.

It should be noted, that the above expressions depend on two scale parameters:
the renormalization scale $\mu$ in strong coupling constant $\alpha_{S}(\mu)$
and the maximum transverse momentum of collinear gluons $Q$ in structure
functions $f(x,Q)$. At low energies the variation of these scale parameters
leads to significant variation of the total cross sections
\cite{Luchinsky:2011wy}, while for the energies about several TeVs the
dependence is not crucial. In our work we will use the value $ Q =\mu=M$, where
$M$ is the mass of the heavy quarkonium. Also we shall neglect the mass
difference between states from same $\chi_J(nP)$ triplet.

It is known, that color-singlet model of $q\bar q$ pair hadronization into
observable meson does not fully describe experimental data. The reason is that
color-singlet is only the first approximation in the Fock structure of
quarkonium state \cite{Bodwin:1994jh}:
\begin{eqnarray}
\notag
|^{2S+1}L_J\rangle &=& O(1) |^{2S+1}L_J{}^{[1]}\rangle \\[1pt]
\notag
&+& O(v) |^{2S+1}(L\pm1)_{J'}{}^{[8]} g\rangle   \qquad\mbox{(E1)}\\[1pt]
\notag
&+& O(v^2) |^{2(S\pm1)+1}L_{J'}{}^{[8]} g\rangle
\qquad\,\,\,\,\mbox{(M1)}\\[1pt]
\notag
&+& O(v^2) |^{2S+1}L_{J}{}^{[1,8]} gg\rangle
\qquad\,\,\,\,\mbox{(E1$\times$E1)}\\[1pt]
\label{eq:octet}
&+& \dots
\end{eqnarray}
where $v$ is the relative velocity of quarks in heavy quarkonium. In the above
expression E1 and M1 are electric ($\Delta L = 1$, $\Delta S = 0$) and magnetic
($\Delta S = 1$, $\Delta L = 0$)  transitions respectively. In the case of
$\chi_c$ mesons equation (\ref{eq:octet}) gives:
\begin{eqnarray}
\notag
|\chi_{cJ}\rangle &=& %
\langle\,\mathcal{O}^{\chi_{cJ}}[^3P_{J}^{[1]}]\,\rangle\,\,|^{3}P_J{}^{[1]}
\rangle 
+\langle\,\mathcal{O}^{\chi_{cJ}}[^3S_1^{[8]}]\,\rangle\,\,|^{3}S_1{}^{[8]}
\rangle+\\[1pt]
\label{eq:octet_chi}
&+&\langle\,\mathcal{O}^{\chi_{cJ}}[^1P_1^{[8]}]\,\rangle\,\,|^{1}P_1{}^{[8]}
\rangle 
+\sum_{J'}\langle\,\mathcal{O}^{\chi_{cJ}}[^3P_{J'}^{[8]}]\,\rangle\,\,|^{3}P_{
J'}{}^{[8]}\rangle +\dots
\end{eqnarray}
Here the fist two terms are order of $O(v)$, and next two terms are of order
$O(v^2)$. 

\begin{figure}[h]
\begin{centering}
\includegraphics[width=\textwidth]{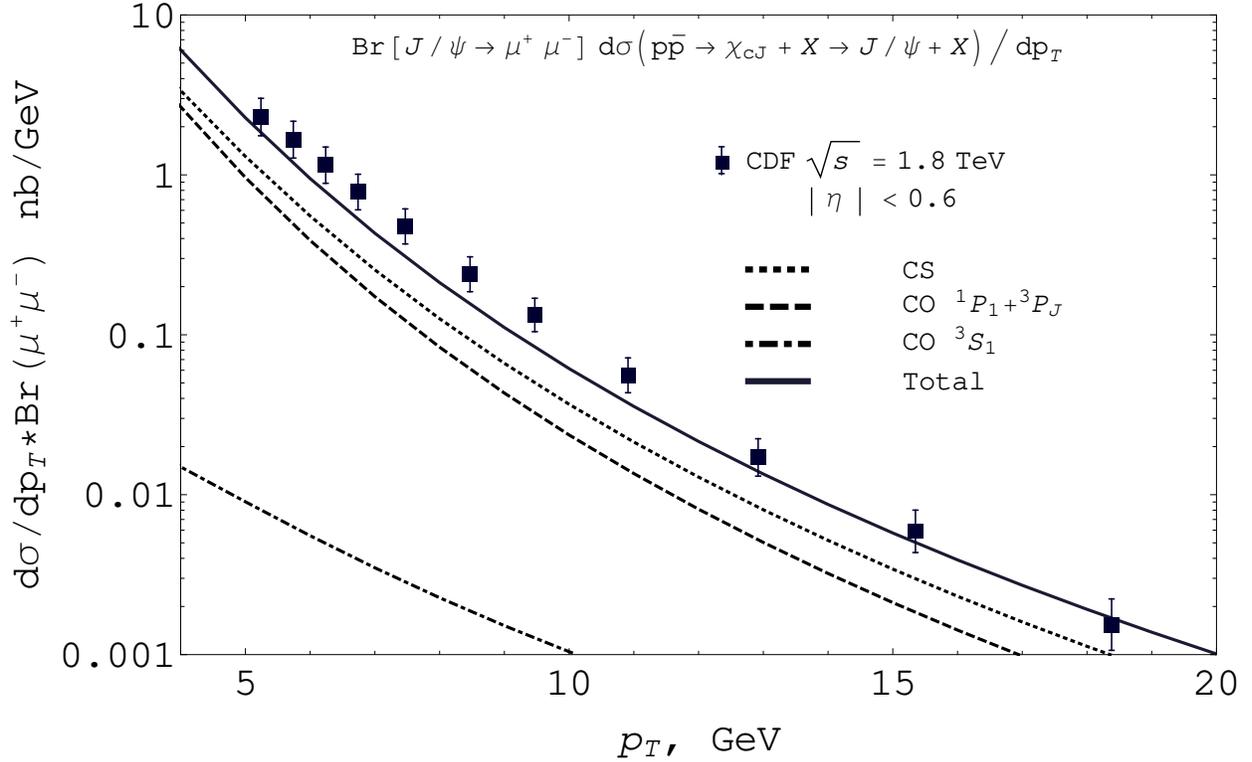}
\par\end{centering}
\caption{Contributions to the $\chi_c$ production from different states. Dotted
line is color singlet contribution. Dashed line is a contribution from
the octet $|^3P_J^{[8]}\rangle$ and $|^1P_1^{[8]}\rangle$ states. Dot-dashed
line is color-octet $|^3S_1^{[8]}\rangle$ state contribution. Solid line is a
sum over all contributions. Experimental points are taken from CDF report
\cite{Abe:1997yz}.
\label{fig:CDF2}}
\end{figure}

In Fig.~\ref{fig:CDF2} the contributions from different states in
(\ref{eq:octet_chi}) to the $p_T$-distributions of $J/\psi$ mesons produced via
$\chi_c$ radiative decays:
\[
\frac{d \sigma}{dp_T}[p\bar p\to\chi_c+X\to J/\psi+X] = 
\Br[\chi_{c1}]\frac{d \sigma}{dp_T}[p\bar p\to\chi_{c1}+X] + 
\Br[\chi_{c2}]\frac{d \sigma}{dp_T}[p\bar p\to\chi_{c2}+X].
\]
are shown in comparison with the CDF experimental data \cite{Abe:1997yz}.
The octet $|^3S_1^{[8]}\rangle$ state gives $p_T$-distribution approximately
described by $\sim 1/p_T^4$, which is far different from the experimental one,
which, in turn, is approximately described by $\sim 1/p_T^6$. So, in order to
fit the experimental data, we should assume, that $|^3S_1^{[8]}\rangle$ gives a
negligibly small contribution in the considered $p_T$ interval and plays role
for higher $p_T$ only. The other states from the Fock expansion
(\ref{eq:octet_chi}), including color-singlet, have the $p_T$-distribution
similar to the experimental measurement. In order to fit the experimental data,
we used CDF data for the $\chi_c$ production and the available data for the
cross section ratio of $\sigma(\chi_{c2})/\sigma(\chi_{c1})$ \cite{LHCb:2012mr,
CDF:2007bra,CMS-PAS-BPH-11-010} (see next section and Fig.~\ref{fig:ratio}). The
singlet contribution is determined
by the singlet matrix element $\mathcal{O}[^3P_{J}^{[1]}]$, which is connected
with the derivative of the radial part of the wave function at the origin:
\[
\langle\mathcal{O}^{\chi_{cJ}}[^3P_{J}^{[1]}]\rangle = \frac{3}{4\pi} (2J+1)
|R'(0)|^2.
\]
For the last one we took the value derived from the $\chi_{c2}$ photonic width
in LO:
\[
|R'(0)|^2 = 0.075\,\mbox{GeV}^5.
\]
The octet matrix elements for $|^3S_1^{[8]}\rangle$ satisfies the multiplicity
relations
\begin{eqnarray*}
\langle\mathcal{O}^{\chi_{cJ}}[^3S_{1}^{[8]}]\rangle &=&
(2J+1)\langle\mathcal{O}^{\chi_{c0}}[^3S_{1}^{[8]}]\rangle\\
\end{eqnarray*}
For the last one we obtained from our fit the following value:
\begin{equation}
\langle\mathcal{O}^{\chi_{c0}}[^3S_{1}^{[8]}]\rangle \lesssim
2\times10^{-5}\,\mbox{GeV}^3
\end{equation}

Since $|^{1}P_1{}^{[8]}\rangle $ and $|^{3}P_J{}^{[8]}\rangle$ states have
similar $p_T$-dependence, they could not be resolved separately and only a
linear combination of the corresponding octet matrix elements can be determined.
In the notations of \cite{Meijer:2007eb} we obtain the following values for the
non-perturbative parametres:
\begin{equation}
\langle R^{\chi_{c0}}[^1P_1^{(8)}] \rangle = \langle
R^{\chi_{c0}}[^3P_{J'}^{(8)}]\rangle = 0.01 \,\mbox{GeV}^5
\end{equation}
The last result is independent on $J'$. Taking into account statistical weight
of the $\chi_{cJ}$-state the corresponding parametres for the $\chi_{cJ}$
production can be recovered:
\begin{eqnarray*}
\langle R^{\chi_{cJ}}[^1P_1^{(8)}] \rangle &=& (2J+1) \langle
R^{\chi_{c0}}[^1P_1^{(8)}]\\
\langle R^{\chi_{cJ}}[^3P_{J'}^{(8)}]\rangle &=&
(2J+1)\langle R^{\chi_{c0}}[^3P_{J'}^{(8)}]\rangle 
\end{eqnarray*}

At this point we need to discusss the results of our fit. In most works (see,
for example, review \cite{Kramer:2001hh}) it is accepted to neglect
the $P$-wave octet contributions due to the velocity power counting rules, and
the following ranges of values of non-perturbative matrix element for the
$|^3S_1^{[8]}\rangle$ state are considered:
\[
\langle\mathcal{O}^{\chi_{c0}}[^3S_{1}^{[8]}]
\rangle \sim
(0.1 - 0.5) \times10^{-2}\,\mbox{GeV}^3
\]
Nevertheless, we state that $|^3S_1^{[8]}\rangle$ contribution gives
inappropriate $p_T$-dependence to describe the experimental data and therefore
octet $P$-wave contributions should be included.

On the other hand, the non-perturbative value of the octet $|^3S_1^{[8]}\rangle$
contribution from work \cite{Kniehl:2006sk} is closer to our result:
\[
\langle\mathcal{O}^{\chi_{c0}}[^3S_{1}^{[8]}]
\rangle \sim
2 \times10^{-4}\,\mbox{GeV}^3,
\]
but the derivative of the radial part of the wave function at the origin is
about several times bigger then ours:
\[
|R'(0)|^2 \sim 0.37\,\mbox{GeV}^5,
\]
so the almost entire contribution is given by the singlet term. We think, that
this value is unreasonably high and does not agree with experimental data.
Possible reason of the contradiction is that authors derive this value from the
photonic width including QCD radiative correction. It is clear, that using
such approach the radiative corrections to the initial state production should
be considered on a par with radiative corrections in the final state, but this
was not performed.

\section{$\chi_b$ production at high $p_T$}
Let us proceed to the $\chi_b$ mesons production. The picture of the bottomonium
production is almost the same as in the charmonium case. The only difference is
that probabilities in Fock expansion (\ref{eq:octet_chi}) have different
values. We shall assume, that relative contributions from different Fock states
are similar to charmonium mesons and do not depend on the radial
excitation number $n$. So, for example, 
\begin{equation}
\label{nonpb1}
\frac{\langle\mathcal{O}[^3P_{J}^{[8]}\to\chi_c]\rangle}%
{\langle\mathcal{O}[^3P_{J}^{[1]}\to\chi_c]\rangle}
=
\frac{\langle\mathcal{O}[^3P_{J}^{[8]}\to\chi_b(nP)]\rangle}%
{\langle\mathcal{O}[^3P_{J}^{[1]}\to\chi_b(nP)]\rangle}
\end{equation}
From the dimension analysis, it is clear, that
\[
\frac{\langle\mathcal{O}[^3S_{1}^{[8]}]\rangle}%
{\langle\mathcal{O}[^3P_{J}^{[1]}]\rangle} \sim \frac{1}{M^2}
\]
So, for $|^3S_1\rangle$ we take 
\begin{equation}
\label{nonpb2}
\frac{\langle\mathcal{O}[^3S_{1}^{[8]}\to\chi_c]\rangle}%
{\langle\mathcal{O}[^3P_{J}^{[1]}\to\chi_c]\rangle}
=
\frac{M_{\chi_b}^2}{M_{\chi_c}^2}\,
\frac{\langle\mathcal{O}[^3S_{1}^{[8]}\to\chi_b(nP)]\rangle}%
{\langle\mathcal{O}[^3P_{J}^{[1]}\to\chi_b(nP)]\rangle}
\end{equation}

\begin{figure}[h]
\begin{centering}
\includegraphics[width=0.8\textwidth]{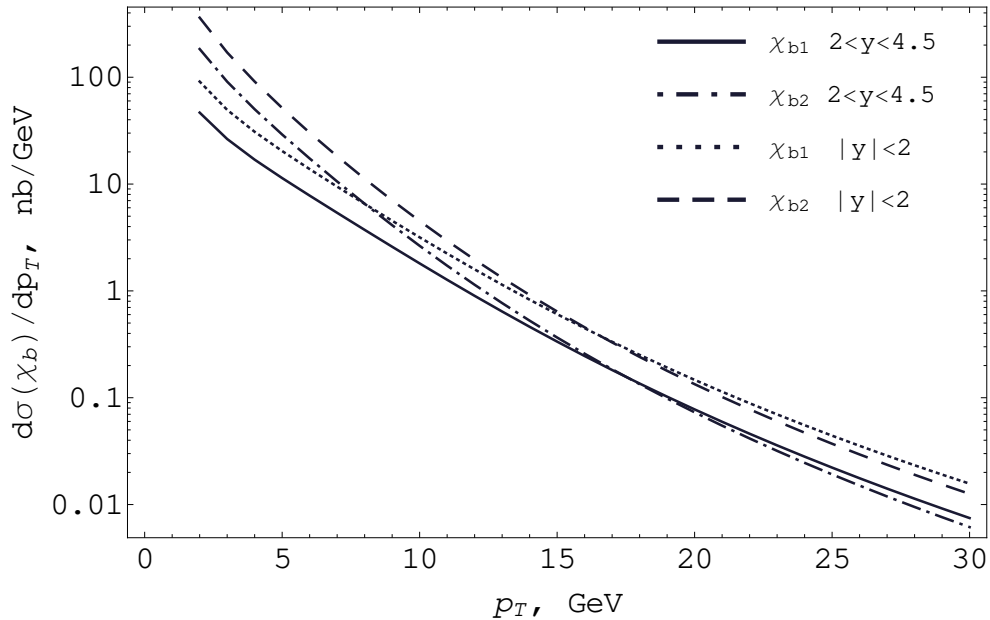}
\par\end{centering}
\caption{Transverse momentum distributions of the $\chi_{b1}$ and $\chi_{b2}$
mesons at $\sqrt{s}=8$ TeV. The derivative of the wave
function at the origin is $|R_n'(0)|^2 = 1\mbox{ GeV}^{5}$, which is in
agreement with potential models, presented in Tab.~\ref{tab:R20}.
Other non-perturbative parametres were obtained in the previous section in the
case of charmonium and corresponds to the total curve in Fig.~\ref{fig:CDF2} and
then evaluated for bottomonium using (\ref{nonpb1}) and (\ref{nonpb2}).
\label{fig:distPTb}}
\end{figure}

In Fig.~\ref{fig:distPTb} we show our results for the transverse momentum
distributions for axial and tensor bottomonium. For the derivative of the wave
function at the origin we take 
\[
|R_n'(0)|^2 = 1\mbox{ GeV}^{5},
\]
which is in agreement with potential models, presented in Tab.~\ref{tab:R20}.
For other non-perturbative parametres we use values obtained in the previous
section and evaluated with (\ref{nonpb1}) and (\ref{nonpb2}).
Presented curves allows one to easily reconstruct the distributions for any
$|R_n'(0)|$ value, by simple multiplying on the corresponding factor from
Tab.~\ref{tab:R20}.
\begin{table}[h]
\begin{centering}
\begin{tabular}{|c|c|c|c|c|c|c|}
\hline 
 & \cite{Munz:1996hb} & \cite{Ebert:2003mu} & \cite{Anisovich:2005jp} &
\cite{Wang:2009er} & \cite{Li:2009nr} & \cite{Hwang:2010iq}\tabularnewline
\hline 
\hline 
$\chi_{b}\left(1P\right)$ & $0.71$ &  $1.02$ &  $2.03$ &  $0.94$ &  $0.84$ & 
$0.51$\tabularnewline
\hline 
$\chi_{b}\left(2P\right)$ & $0.99$ &  $0.88$  &  $2.2$ &  $1.13$ &  $0.98$ & 
$0.25$\tabularnewline
\hline 
$\chi_{b}\left(3P\right)$ & --- & --- &  $2.43$ &  $1.17$ &  $1.03$ &
---\tabularnewline
\hline 
\end{tabular}
\par\end{centering}
\caption{$\left|R'_{nP}\left(0\right)\right|^{2}$ (in $\GeV^{5}$)  
from different potential models\label{tab:R20}}
\end{table}

\begin{figure}[h]
\begin{centering}
\includegraphics[width=0.8\textwidth]{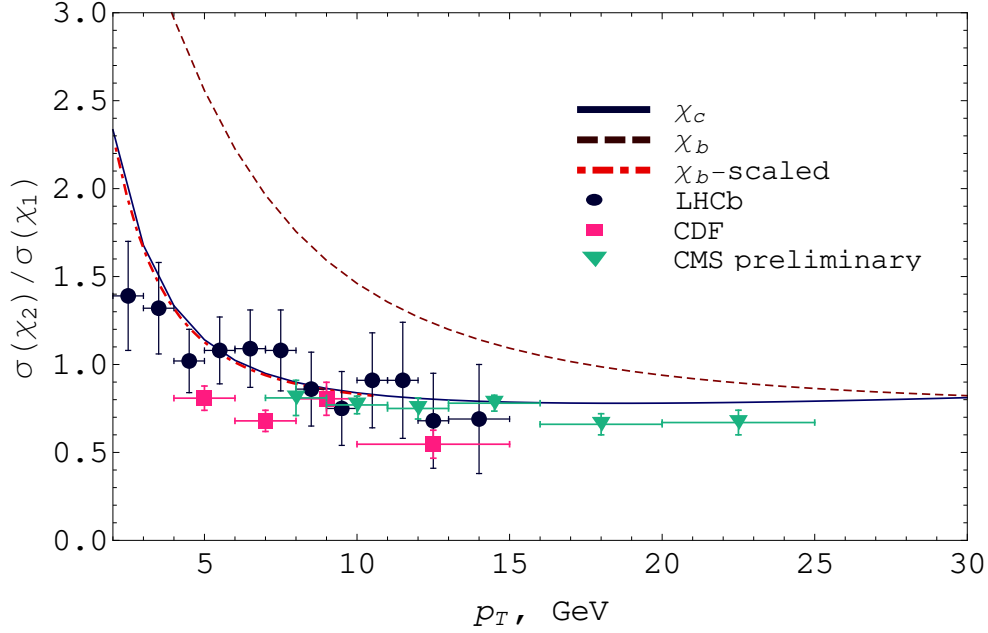}
\par\end{centering}
\caption{Transverse momentum distributions of the
$d\sigma\left[\chi_{2}\right]/d\sigma[\chi_{1}]$
ratio. Solid and dashed lines stand for charmonium and bottomonium
mesons. The dot-dashed line corresponds to the rescaled bottomonium ratio:
$\sigma_{b2}/\sigma_{b1}(M_{\chi_c}/M_{\chi)b}\, p_T)$. As it is seen, it
almost matches the charmonium curve. The experimental results for charmonium
from LHC \cite{LHCb:2012mr} are shown with dots, CDF \cite{CDF:2007bra} --- with
rectangles, and CMS \cite{CMS-PAS-BPH-11-010} --- with triangles.}
\label{fig:ratio}
\end{figure}

In Fig.~\ref{fig:ratio} the $p_T$-dependence of the $\chi_Q(J=2)/\chi_Q(J=1)$
ratio is depicted. In case of charmonium mesons we compare our
results with the LHCb data \cite{LHCb:2012mr}, CDF data \cite{CDF:2007bra} and
CMS preliminary \cite{CMS-PAS-BPH-11-010}. Both for $(c\bar c)$ and $(b\bar b)$
mesons we present the cross section ratio in the rapidity range $2<y<4.5$, but
it is clear that this ratio depends weakly on the rapidity cut, so these
predictions should be  also valid for another rapidity regions. One can easily
see, that within the experimental errors our results for charmonia agree well
with experiment, so approach used in our paper is valid. It is clear, that
since the cross section is proportional to the $|R'(0)|^2$ and if the mass gap
between $\chi_{J}\left(nP\right)$ states with different radial numbers $n$
is neglected, the ratio $d\sigma[\chi_{2}(nP)]/d\sigma[\chi_{1}(nP)]$
does not depend on $n$ and curves for charmonia and bottomonia presented on
Fig.~\ref{fig:ratio} are universal.

One interesting property can be seen from Fig.~\ref{fig:ratio}: the
bottomonium curve matches the charmonium curve if we perform the rescaling of
the $p_T$ variable: $p_T\to (M_{\chi_c}/M_{\chi_b}) p_T \approx (1/3)p_T$.
Moreover, this fact is exact model-independent theoretical result, which can be
obtained from the dimension analysis. Indeed, the cross section of the quarkonia
production depends on three dimensional parameters: hadronic energy, transverse
momentum of final quarkonium and its mass:
\[
\frac{d \sigma_J}{d p_T} \equiv \frac{d \sigma_J}{d p_T}(s,p_T,M)
\]
The cross section ratio $J=2$ and $J=1$ is dimensionless function of these
parameters. Hence, we can write:
\begin{equation}
\label{eq:scaleRatio}
\left. \frac{d\sigma_{2}(z^2 s,z p_T, zM)}{dp_T}\right/\frac{d\sigma_{1}(z^2
s,zp_T,zM)}{dp_T}
= 
\left. \frac{d\sigma_{2}(s, p_T,
M)}{dp_T}\right/\frac{d\sigma_{1}(s,p_T,M)}{dp_T}
\end{equation}
Taking $M=M_{\chi_c}$ and $z=M_{\chi_b}/M_{\chi_c}$, we obtain
\begin{equation}
\label{eq:scaleRatio}
\left. \frac{d\sigma_{b2}(zp_T;s)}{dp_T}\right/\frac{d\sigma_{b1}(zp_T;s)}{dp_T}
= 
\left.
\frac{d\sigma_{c2}(p_T;s/z^2)}{dp_T}\right/\frac{d\sigma_{c1}(p_T;s/z^2)}{dp_T}.
\end{equation}
Hence, if we know the charmonium ratio transverse momentum distribution at some
c.m. energy $s$, we can easily reconstruct the bottomonium ratio distribution at
energy $s\, (M_{\Upsilon}/M_{J/\psi})^2$. This relation is not violated by
the scale dependence of $\alpha_s$ and PDFs, since they are canceled in the
ratio.

It is intuitively clear, that at least for high energies the ratio weakly
depends on $s$, since both axial and tensor meson acquire
universal behaviour in $s$, so the $s$-dependence cancels in this ratio. This
argument does not work for low $s$, when another partonic channels becomes
significant. The value of $z$ is about \( z \sim 3\),
so the bottomonium distribution at $\sqrt{s} = 8$ TeV corresponds to the
charmonium distribution at $\sqrt{s} \sim 2.6$ TeV, which is still very high, so
the gluonic subprocess dominates. We can expect, that (\ref{eq:scaleRatio})
can be approximately rewritten for high energies:
\begin{equation}
\label{eq:scaleRatioApprox}
\left. \frac{d\sigma_{b2}(zp_T;s)}{dp_T}\right/\frac{d\sigma_{b1}(zp_T;s)}{dp_T}
= 
\left. \frac{d\sigma_{c2}(p_T;s)}{dp_T}\right/\frac{d\sigma_{c1}(p_T;s)}{dp_T}.
\end{equation}

\subsection{Radial excitations}

Let us now discuss the production of excited $\chi_{b}$-mesons at
LHC. Recently $\chi_{b}(1P)$-, $\chi_{b}(2P)-$ and $\chi_{b}(3P)$-mesons
were observed in $\Upsilon(1S)\gamma$ and $\Upsilon(2S)\gamma$ modes
by ATLAS Collaboration \cite{Aad:2011ih} and D0 \cite{D0:2012gh} Collaboration.
In these experiments the radial excitation of the P-wave $\chi_{b}$ states was
observed in the radiative transitions to the S-wave $\Upsilon$ states.
Unfortunately the exact values of the cross sections were not measured yet, but
we can reconstruct them theoretically. 

The experimentally observable quantities are:
\begin{eqnarray}
\sigma^{\Th}\left[nP,mS\right] & = &
\sigma^{\Th}\left[pp\to\chi_{b}
\left(1P\right)+X\to\Upsilon\left(mS\right)\gamma+X\right]=\sum_{J=0}^{2}\Br_{J}
\left[nP,mS\right]\sigma_{J}^{\Th}\left[nP\right],\label{eq:sigmaN}
\end{eqnarray}
where the following notations were used:
\begin{eqnarray*}
\sigma_{J}^{\Th}\left[nP\right] & = &
\sigma^{\Th}\left[pp\to\chi_{bJ}\left(nP\right)+X\right],\\
\Br_{J}\left[nP,mS\right] & = &
\Br\left[\chi_{bJ}\left(nP\right)\to\Upsilon\left(1S\right)\gamma\right].
\end{eqnarray*}
The branching fractions of radiative $\chi_{b}(1P)$- and
$\chi_{b}(2P)$-mesons decays are known experimentally \cite{PDBook}: 
\begin{align*}
\Br_{1}\left[1P,1S\right] & = 35\% \pm 8\%,&\Br_{2}\left[1P,1S\right] &= 22\%\pm
4\%,\\
\Br_{1}\left[2P,1S\right] & = 8.5\%\pm 1.3\%,&\Br_{2}\left[2P,1S\right] &=
7.1\%\pm 1\%,\\
\Br_{1}\left[2P,2S\right] & = 21\%\pm 4\%,&\Br_{2}\left[2P,2S\right] &= 16\%\pm
2.4\%.
\end{align*}
In the case of scalar bottomonium the branching fractions are small,
so we do not include it in the sum (\ref{eq:sigmaN}). 

In order to calculate the total cross sections, we should integrate the cross
sections distributions obtained in the previous section and presented on
Fig.~\ref{fig:distPTb} over appropriate $p_T$ region: $\Delta < p_T <
(s-M^2)/2\sqrt{s}$. We use experimental values $\Delta = 12$ GeV for
ATLAS ($|y|<2$) \cite{Aad:2011ih}, and $\Delta=6$ GeV for LHCb
($2<y<4.5$) \cite{LHCb-CONF-2012-020}. Using these values we obtained the
following prediction for the cross sections:
\begin{align*}
 \mbox{(LHCb)}&&  \frac{1}{|R_n'(0)|^2}\sigma^{\Th}[\chi_{b1}(nP)]& =
34.4\frac{\mathrm{nb}}{\mathrm{GeV}^5} %
 & \frac{1}{|R_n'(0)|^2}\sigma^{\Th}[\chi_{b2}(nP)]& = 
43\frac{\mathrm{nb}}{\mathrm{GeV}^5}\\
 \mbox{(ATLAS)}&& \frac{1}{|R_n'(0)|^2}\sigma^{\Th}[\chi_{b1}(nP)]& = 
5.2\frac{\mathrm{nb}}{\mathrm{GeV}^5}%
 & \frac{1}{|R_n'(0)|^2}\sigma^{\Th}[\chi_{b2}(nP)]& = 
5.6\frac{\mathrm{nb}}{\mathrm{GeV}^5}
\end{align*}
and for the ratio $\sigma[\chi_{b}(2P)]/\sigma[\chi_{b}(1P)]$ in both cases:
\begin{eqnarray*}
\frac{\sigma^{\Th}\left[2P,1S\right]}{\sigma^{\Th}\left[1P,1S\right]} & = &
\left(0.29 {}\pm{} 0.01^{th} {}\pm{}
0.1^{br}\right){}\left|\frac{R'_{2P}}{R'_{1P}}\right|^{2},
\end{eqnarray*}
Here the first error is the error of our theoretical method, while the second 
is the uncertainty in the experimental values of branchings fractions. This
ratio is determined by the ratio of $2P$- and $1P$-states wave function
derivatives at the origin (see Table \ref{tab:R20}). In Table \ref{tab22} we
show the predictions of different potential models for this ratio.

\begin{table}[h]
\begin{centering} 
\begin{tabular}{|c|ccccc|}
\hline 
Potential model & \multicolumn{5}{c|}{Ratio
$\chi_b(2P)/\chi_b(1P)$}\tabularnewline
\hline
\hline
\cite{Munz:1996hb}       & $0.4$
&$\pm$&$0.01^{th}$&$\pm$&$0.14^{br}$\tabularnewline
\cite{Ebert:2003mu}      &
$0.25$&$\pm$&$0.01^{th}$&$\pm$&$0.1^{br}$\tabularnewline
 \cite{Anisovich:2005jp} &
$0.32$&$\pm$&$0.02^{th}$&$\pm$&$0.1^{br}$\tabularnewline
\cite{Wang:2009er}       &
$0.34$&$\pm$&$0.01^{th}$&$\pm$&$0.12^{br}$\tabularnewline
 \cite{Li:2009nr}        &
$0.34$&$\pm$&$0.01^{th}$&$\pm$&$0.12^{br}$\tabularnewline
\cite{Hwang:2010iq}      & $0.14$&     &           &$\pm$&$0.05^{br}$
\tabularnewline
\hline 
\end{tabular}
\par\end{centering}
\caption{Theoretical predictions for the ratio of $\chi_{b}(2P)$ and
$\chi_{b}(1P)$
production cross sections in different potential models\label{tab22} }
\end{table}

In the case of $\chi_{b}(3P)$-mesons production the corresponding
branching fractions to $\Upsilon(1S)$ are not known yet, so we cannot
perform similar analysis. However, we can estimate the radiative branching
fraction of tensor meson using the following assumptions. We assume, that the
main hadronic decay channel of tensor meson is 2-gluon decay:
\begin{equation*}
\Gamma[\chi_{b2}(3P)\to gg] = \frac{128}{5}\alpha_S^2 \frac{|R_3'(0)|^2}{M^4}
\end{equation*}
 and the radiative branching fraction of the $\chi_{b2}(3P)$ is equal to
\begin{equation*}
\Br_{2}\left[3P,1S\right] = \frac{\Gamma[\chi_{b2}(3P) \to
\Upsilon(1S)\gamma]}{\Gamma[\chi_{b2}(3P) \to \Upsilon(1,2,3S)\gamma] +
\Gamma[\chi_{b2}(3P)\to gg]}
\end{equation*}
The radiative width can be obtained from the potential models
\cite{Anisovich:2005jp,Li:2009nr}. Using the cross section of $\Upsilon(1S)$
produced via $\chi_{b1,2}(3P)$ radiative decays
\begin{eqnarray*}
\sigma^{\Th}\left[3P,1S\right] & = &
\Br_{1}\left[3P,1S\right]\sigma_{1}^{\Th}\left[3P\right]+\Br_{2}\left[3P,
1S\right]\sigma_{2}^{\Th}\left[3P\right],
\end{eqnarray*}
we can express the unknown branching fraction of the
$\chi_{b1}(3P)\to\Upsilon(1S)\gamma$ decay through the ratio of the $3P$ and
$1P$ production cross sections. The results are presented in Table
\ref{tab:Br31}, where we denoted
\[
\gamma =  \frac{\sigma^{\Th}\left[3P,1S\right]}{\sigma^{\Th}\left[1P,1S\right]}
\]
So, if the experimental value of $\gamma$ is found, the estimation on the
radiative width $\chi_b1(3P)\to\Upsilon(1S)\gamma$ will be found too.  Similar
estimations can be also performed for the $\chi_b\to\Upsilon(2S)\gamma$
processes.
\begin{table}
\begin{centering} 
\begin{tabular}{|c|c|c|}
\hline 
Potential model & $\Br_{2}\left[3P,1S\right]$ &
$\Br_{1}\left[3P,1S\right]$\tabularnewline
\hline
\hline
\cite{Anisovich:2005jp} & $1.7\%$ & $\left((41 \pm
9)\,\gamma-1.3\right)$\%\tabularnewline
\cite{Li:2009nr} & $5.4\%$ & $\left((42 \pm
9)\,\gamma-4.1\right)$\%\tabularnewline
\hline
\end{tabular}
\par\end{centering}
\caption{Branching fractions of $\chi_b(3P)\to\Upsilon(1S)\gamma$
decays\label{tab:Br31}}
\end{table}

\section{Conclusions}
The present paper is devoted to the $\chi_b$-meson production in hadronic
experiments. Recently ATLAS \cite{Aad:2011ih} and D0 \cite{D0:2012gh}
Collaborations reported about observation of $\chi_b(1,2,3P)$ states, so
theoretical description of these processes is desirable.

In our paper we use both color singlet and color octet models for description of
the $P$-wave heavy quarkonia production. For high energy reactions these mesons
are produced mainly in gluon-gluon interactions and the cross sections of the
corresponding processes can be written as a convolution of partonic cross
sections and gluon distribution functions in initial hadrons. If one uses usual
distribution functions integrated over the gluon transverse momentum and works
at leading order (i.e. only $gg\to\chi_b$ partonic reactions are considered),
the information about the transverse momentum distribution of final quarkonia
$p_T$ is lost. It is clear, that in order to describe this distribution one has
to use next to leading order results, when additional gluons in the final state
are present. For high $p_T$ values processes with emission of additional gluons
are suppressed by small strong coupling constant, so main contributions come
from reactions with single hard gluon in the final state. For this reason we
consider NLO partonic reactions $gg\to\chi_{QJ} g$. It should be noted, that in
this approach the production of axial $\chi_1$-meson is possible, while at
leading order it is forbidden by Landau-Yang theorem.

In our article we discuss transverse momentum distributions of
$\chi_{c,b}$ mesons. It is shown, that in the case of $\chi_c$-meson production
our predictions agree well with the available experimental data. In case of
$\chi_b(nP)$ mesons, we give predictions of their $p_T$-distributions and
absolute cross sections  in ATLAS and LHCb experiments. One interesting property
we found is that, according to our estimations, distributions of the
ratios $\left.[d\sigma(\chi_2)/dp_T]\right/[d\sigma(\chi_1)/dp_T]$ for
charmonium and bottomonium mesons coincide if the scaling $p_T^{\chi_b} \to
(M_{\chi_c}/M_{\chi_b})p_T^{\chi_b}$ is performed. Using
existing information
about radial excitation of $\chi_b$-mesons from different potential models we
predict the ratio of $\chi_b(2P)$ and $\chi_b(1P)$ yields. Unfortunately, ATLAS
and D0 Collaborations do not provide the information about the normalization of
their results, so currently comparison with experimental data is not possible.
As for newly observed $\chi_b(3P)$ meson state, we show, that from the ratios of
$\chi_b(3P)$ and $\chi_b(1,2P)$ production cross sections one can determine the
branching fractions of $\chi_{b1}(3P)\to\Upsilon(1,2S)\gamma$ decays.

\section*{Acknowledgements}
We would like to thank Dr. Vakhtang Kartvelishvili, Vladimir Obraztsov and
Alexey Novoselov for useful criticism. We also want to thank colleagues from the
LHCb collaboration and specially Ivan Belayev for discussions.

The work was financially supported by Russian Foundation for Basic
Research (grant \#10-02-00061a) and the grant of the president of
Russian Federation (grant \#MK-3513.2012.2).

\bibliographystyle{utphys}
\bibliography{literature}

\providecommand{\href}[2]{#2}\begingroup\raggedright\begin{thebibliography}{10}

\bibitem{Aad:2011ih}
{\bfseries ATLAS} Collaboration, G.~Aad {\em et~al.}, ``{Observation of a new
  $\chi_b$ state in radiative transitions to $\Upsilon(1S)$ and $\Upsilon(2S)$
  at ATLAS},''
\href{http://arxiv.org/abs/1112.5154}{{\ttfamily arXiv:1112.5154 [hep-ex]}}.

\bibitem{D0:2012gh}
{\bfseries D0} Collaboration, V.~M. Abazov {\em et~al.}, ``{Observation of a
  narrow mass state decaying into Upsilon(1S) + gamma in ppbar collisions at
  sqrt(s) = 1.96 TeV},'' \href{http://arxiv.org/abs/1203.6034}{{\ttfamily
  arXiv:1203.6034 [hep-ex]}}.

\bibitem{Kartvelishvili:1978id}
V.~Kartvelishvili, A.~Likhoded, and S.~Slabospitsky, ``{D meson and Psi meson
  production in hadronic interactions},''
{\em Sov.J.Nucl.Phys.} {\bfseries 28} (1978) 678.

\bibitem{Gershtein:1980jb}
S.~Gershtein, A.~Likhoded, and S.~a. Slabospitsky, ``{Charmed particle
  inclusive spectra in photoproduction processes},''
{\em Sov.J.Nucl.Phys.} {\bfseries 34} (1981) 128.

\bibitem{Gribov:1984tu}
L.~Gribov, E.~Levin, and M.~Ryskin, ``{Semihard Processes in QCD},''
\href{http://dx.doi.org/10.1016/0370-1573(83)90022-4}{{\em Phys.Rept.}
  {\bfseries 100} (1983) 1--150}.

\bibitem{Catani:1990eg}
S.~Catani, M.~Ciafaloni, and F.~Hautmann, ``{High-energy factorization and
  small x heavy flavor production},''
\href{http://dx.doi.org/10.1016/0550-3213(91)90055-3}{{\em Nucl.Phys.}
  {\bfseries B366} (1991) 135--188}.

\bibitem{Collins:1991ty}
J.~C. Collins and R.~Ellis, ``{Heavy quark production in very high-energy
  hadron collisions},''
\href{http://dx.doi.org/10.1016/0550-3213(91)90288-9}{{\em Nucl.Phys.}
  {\bfseries B360} (1991) 3--30}.

\bibitem{Kniehl:2006sk}
 B.A.~Kniehl, D.V.~Vasin, and V.A.~Saleev,  ``{Charmonium production at high energy in
  the $k_{T}$ -factorization approach},''
  \href{http://dx.doi.org/10.1103/PhysRevD.73.074022}{{\em Phys.Rev.}
  {\bfseries D73} (2006) 074022},
\href{http://arxiv.org/abs/hep-ph/0602179}{{\ttfamily arXiv:hep-ph/0602179
  [hep-ph]}}.

\bibitem{Abe:1997yz}
{\bfseries CDF} Collaboration, F.~Abe {\em et~al.}, ``{Production of $J/\psi$
  mesons from $\chi_c$ meson decays in $p\bar{p}$ collisions at $\sqrt{s} =
  1.8$ TeV},''
{\em Phys.Rev.Lett.} {\bfseries 79} (1997) 578--583.

\bibitem{LHCb:2012mr}
{\bfseries LHCb} Collaboration,  {\em et~al.}, ``{Measurement of the
  cross-section ratio $\sigma(\chi_{c2})/\sigma(\chi_{c1})$ for prompt $\chi_c$
  production at $\sqrt{s}=7$ TeV},''
\href{http://arxiv.org/abs/1202.1080}{{\ttfamily arXiv:1202.1080 [hep-ex]}}.

\bibitem{Gastmans:1987be}
R.~Gastmans, W.~Troost, and T.~T. Wu, ``{Production of heavy quarkonia from
  gluons},''
\href{http://dx.doi.org/10.1016/0550-3213(87)90493-7}{{\em Nucl.Phys.}
  {\bfseries B291} (1987) 731}.

\bibitem{Cho:1995ce}
P.~L. Cho and A.~K. Leibovich, ``{Color octet quarkonia production. 2.},''
  \href{http://dx.doi.org/10.1103/PhysRevD.53.6203}{{\em Phys.Rev.} {\bfseries
  D53} (1996) 6203--6217},
\href{http://arxiv.org/abs/hep-ph/9511315}{{\ttfamily arXiv:hep-ph/9511315
  [hep-ph]}}.

\bibitem{Klasen:2003zn}
M.~Klasen, B.A.~Kniehl, L.N.~Mihaila, and M.~Steinhauser, ``{Charmonium production
  in polarized high-energy collisions},''
  \href{http://dx.doi.org/10.1103/PhysRevD.68.034017}{{\em Phys.Rev.}
  {\bfseries D68} (2003) 034017},
\href{http://arxiv.org/abs/hep-ph/0306080}{{\ttfamily arXiv:hep-ph/0306080
  [hep-ph]}}.

\bibitem{Meijer:2007eb}
M.M.~Meijer, J.~Smith, and W.L.~van Neerven, ``{Helicity amplitudes for charmonium
  production in hadron-hadron and photon-hadron collisions},''
  \href{http://dx.doi.org/10.1103/PhysRevD.77.034014}{{\em Phys.Rev.}
  {\bfseries D77} (2008) 034014},
\href{http://arxiv.org/abs/0710.3090}{{\ttfamily arXiv:0710.3090 [hep-ph]}}.

\bibitem{Luchinsky:2011wy}
A.~Luchinsky and S.~Poslavsky, ``{Inclusive charmonium production at PANDA
  experiment},'' \href{http://dx.doi.org/10.1103/PhysRevD.85.074016}{{\em Phys.
  Rev. D} {\bfseries 85} (Apr, 2012) 074016},
\href{http://arxiv.org/abs/1110.4989}{{\ttfamily arXiv:1110.4989 [hep-ph]}}.

\bibitem{Bodwin:1994jh}
G.~T. Bodwin, E.~Braaten, and G.~P. Lepage, ``{Rigorous QCD analysis of
  inclusive annihilation and production of heavy quarkonium},''
  \href{http://dx.doi.org/10.1103/PhysRevD.51.1125, 10.1103/PhysRevD.55.5853,
  10.1103/PhysRevD.51.1125, 10.1103/PhysRevD.55.5853}{{\em Phys.Rev.}
  {\bfseries D51} (1995) 1125--1171},
\href{http://arxiv.org/abs/hep-ph/9407339}{{\ttfamily arXiv:hep-ph/9407339
  [hep-ph]}}.

\bibitem{Kramer:2001hh}
.~Kramer, Michael, ``{Quarkonium production at high-energy colliders},''
  \href{http://dx.doi.org/10.1016/S0146-6410(01)00154-5}{{\em
  Prog.Part.Nucl.Phys.} {\bfseries 47} (2001) 141--201},
\href{http://arxiv.org/abs/hep-ph/0106120}{{\ttfamily arXiv:hep-ph/0106120
  [hep-ph]}}.

\bibitem{Munz:1996hb}
C.~R. Munz, ``{Two photon decays of mesons in a relativistic quark model},''
  \href{http://dx.doi.org/10.1016/S0375-9474(96)00265-5}{{\em Nucl.Phys.}
  {\bfseries A609} (1996) 364--376},
\href{http://arxiv.org/abs/hep-ph/9601206}{{\ttfamily arXiv:hep-ph/9601206
  [hep-ph]}}.

\bibitem{Ebert:2003mu}
D.~Ebert, R.~Faustov, and V.~Galkin, ``{Two photon decay rates of heavy
  quarkonia in the relativistic quark model},''
  \href{http://dx.doi.org/10.1142/S021773230300971X}{{\em Mod.Phys.Lett.}
  {\bfseries A18} (2003) 601--608},
\href{http://arxiv.org/abs/hep-ph/0302044}{{\ttfamily arXiv:hep-ph/0302044
  [hep-ph]}}.

\bibitem{Anisovich:2005jp}
V.~Anisovich, L.~Dakhno, M.~Matveev, V.~Nikonov, and A.~Sarantsev,
  ``{Quark-antiquark states and their radiative transitions in terms of the
  spectral integral equation. I. Bottomonia},''
  \href{http://dx.doi.org/10.1134/S1063778807010097}{{\em Phys.Atom.Nucl.}
  {\bfseries 70} (2007) 63--92},
\href{http://arxiv.org/abs/hep-ph/0510410}{{\ttfamily arXiv:hep-ph/0510410
  [hep-ph]}}.

\bibitem{Wang:2009er}
G.-L. Wang, ``{Annihilation Rate of 2++ Charmonium and Bottomonium},''
  \href{http://dx.doi.org/10.1016/j.physletb.2009.03.030}{{\em Phys.Lett.}
  {\bfseries B674} (2009) 172--175},
\href{http://arxiv.org/abs/0904.1604}{{\ttfamily arXiv:0904.1604 [hep-ph]}}.

\bibitem{Li:2009nr}
B.-Q. Li and K.-T. Chao, ``{Bottomonium Spectrum with Screened Potential},''
  \href{http://dx.doi.org/10.1088/0253-6102/52/4/20}{{\em Commun.Theor.Phys.}
  {\bfseries 52} (2009) 653--661},
\href{http://arxiv.org/abs/0909.1369}{{\ttfamily arXiv:0909.1369 [hep-ph]}}.

\bibitem{Hwang:2010iq}
C.-W. Hwang and R.-S. Guo, ``{Two-photon and two-gluon decays of p-wave heavy
  quarkonium using a covariant light-front approach},''
  \href{http://dx.doi.org/10.1103/PhysRevD.82.034021}{{\em Phys.Rev.}
  {\bfseries D82} (2010) 034021},
\href{http://arxiv.org/abs/1005.2811}{{\ttfamily arXiv:1005.2811 [hep-ph]}}.

\bibitem{CDF:2007bra}
{\bfseries CDF} Collaboration, A.~Abulencia {\em et~al.}, ``{Measurement of
  $\sigma(\chi(c2)B(\chi(c2) \to J/\psi\gamma) / \sigma(\chi(c1)B(\chi(c1) \to
  J/\psi\gamma)$ in p anti-p collisions at $\sqrt{s} = 1.96$-TeV},''
  \href{http://dx.doi.org/10.1103/PhysRevLett.98.232001}{{\em Phys. Rev. Lett.}
  {\bfseries 98} (2007) 232001},
\href{http://arxiv.org/abs/hep-ex/0703028}{{\ttfamily arXiv:hep-ex/0703028}}.

\bibitem{CMS-PAS-BPH-11-010}
{\bfseries CMS} Collaboration, ``Measurement of sigma(chic2)/ sigma(chic1) at
  sqrt(s) = 7 tev,''. CMS-PAS-BPH-11-010.

\bibitem{PDBook}
{K. Nakamura et al. (Particle Data Group)}, ``Review of particle physics,''
  \href{http://dx.doi.org/doi:10.1088/0954-3899/37/7A/075021}{{\em Journal of
  Physics G} {\bfseries 37} (2010) 075021}. \url{http://pdg.lbl.gov}.

\bibitem{LHCb-CONF-2012-020}
``{Observation of the $\chi_b(3P)$ state at LHCb in $pp$ collisions at
  $\sqrt{s}=7$~TeV},''.
  \href{http://cdsweb.cern.ch/record/1456078?ln=en}{LHCb-ANA-2012-031}.

\end{thebibliography}\endgroup
\end{document}